\documentclass[a4paper]{article}
\usepackage{multirow}
\usepackage{multicol}
\usepackage{diagbox}
\usepackage{CJK}
\usepackage{INTERSPEECH2019}

\title{The FFSVC 2020 Evaluation Plan}

 \name{Xiaoyi Qin$^{1}$, Ming Li$^{1}$, Hui Bu$^{4}$, Rohan Kumar Das$^{2}$, Wei Rao$^{2}$, Shrikanth Narayanan$^{3}$, Haizhou Li$^{2}$}

\address{$^{1}$Data Science Research Center, Duke Kunshan University, Kunshan, China \\
			$^{2}$Department of Electrical \& Computer Engineering, National University of Singapore, Singapore \\
			$^{3}$Signal Analysis and Interpretation Lab, University of Southern California, Los Angeles, USA \\
			 $^{4}$AI Shell Foundation, Beijing, China}
\email{}
\begin{document}

\maketitle

\section{Introduction}
Speaker verification is a key technology in speech processing and biometric authentication, which has broad impact on our daily lives, e.g. security, customer service, mobile devices, smart speakers. Recently, speech based human computer interaction has become more and more popular in far-field smart home and smart city applications, e.g. mobile devices, smart speakers, smart TVs, automobiles. Due to the usage of deep learning methods, the performances of speaker verification in telephone channel and close-talking microphone channel have been enhanced dramatically. However, there are still some open research questions that can be further explored for speaker verification in the far-field and complex environments, including but not limited to

\begin{itemize}
\item{Far-field text-dependent speaker verification for wake up control}
\item{Far-field text-independent speaker verification with complex environments}
\item{Far-field speaker verification with cross-channel enrollment and test}
\item{Far-field speaker verification with single multi-channel microphone array}
\item{Far-field speaker verification with multiple distributed microphone arrays}
\item{Far-field speaker verification with front-end speech enhancement methods}
\item{Far-field speaker verification with end-to-end modeling using data augmentation}
\item{Far-field speaker verification with front-end and back-end joint modeling}
\item{Far-field speaker verification with transfer learning and domain adaptation}
\end{itemize}

The Far-Field Speaker Verification Challenge 2020 (FFSVC20)  is designed to boost the speaker verification research with special focus on far-field distributed microphone arrays under noisy conditions in real scenarios. The objectives of this challenge are to: 1) benchmark the current speech verification technology under this challenging condition, 2) promote the development of new ideas and technologies in speaker verification, 3) provide an open, free, and large scale speech database to the community that exhibits the far-field characteristics in real scenes.

The challenge has three tasks in different scenes.
\begin{itemize}

\item{Task 1: Far-Field Text-Dependent Speaker Verification from single microphone array}

\item{Task 2: Far-Field Text-Independent Speaker Verification from single microphone array}

\item{Task 3: Far-Field Text-Dependent Speaker Verification from distributed microphone arrays}

\end{itemize}

All three tasks follow the cross-channel setup. The recordings of close-talking cellphone will be selected as enrollment and the recordings of far-field microphone array will be used for test.

\section{Database}

\begin{table}[t]
  \caption{The details of the FFSVC20 challenge data}
  \label{tab:data details}
  \centering
  \begin{tabular}[c]{lll}
    \toprule
  Utterence ID&  Content & Noise \\
    \midrule
  \multirow{2}*{001-030}&  ni hao, mi ya & F - TV/Office + electric fan  \\
  &(text-dependent) & T - electric fan \\
  091-  & text independent & S - clean \\
  \bottomrule
  \end{tabular}
\end{table}

\subsection{The FFSVC20 Challenge Database}

\subsubsection{The DMASH Database}
\begin{figure}[t]
  \centering
  \includegraphics[width=0.9\linewidth]{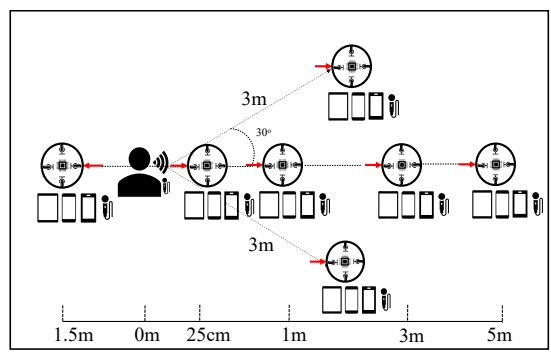}
  \caption{The  setup of the recording environment}
  \label{fig:f1}
\end{figure}

The Distributed Microphone Arrays in Smart Home (DMASH) database is recorded in real smart home scenarios with two different rooms. The detail information of room sizes will be released after the challenge.  Figure \ref{fig:f1} shows the recording environment setup of DMASH, includes one close-talking microphone;  one iPhone, one Android phone, one iPad, one microphone and one circular microphone array (named PCM) placed at 25cm, 1m,3m,5m, left 3m, right 3m and -1.5m distances, respectively.

\subsubsection{FFSVC20 Challenge Data Description}
The FFSVC20 challenge database is part of the DMASH Database. The recording devices include one close-talking microphone (48kHz, 16 bit), one iPhone (48kHz, 16 bit) at 25cm distance and 6 circular microphone arrays (16kHz, 16bit, 16 microphones, 5cm radius). The language is Mandarin. Text content include {\textit{ni hao mi ya}} as text dependent utterances as well as other text independent ones.

\begin{figure}[t]
  \centering
  \includegraphics[width=0.85\linewidth]{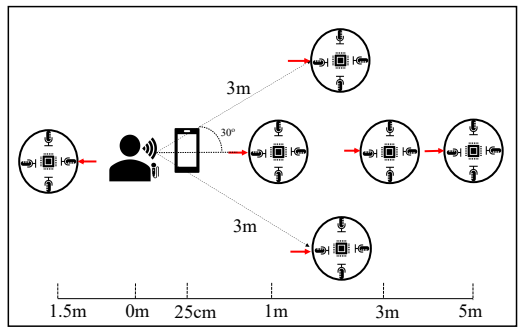}
  \caption{The  setup of the FFSVC20 challenge data}
  \label{fig:f2}
\end{figure}

The data collection setup of the challenge database is shown in  Figure \ref{fig:f2}. Red arrow points to channel 0 of microphone arrays. Each speaker visits 3 times with 7-15 days gap.

The first letter of the file name denotes the visit index. \textit{F} stands for the speaker's first visit, \textit{S} denotes the speaker's second visit and \textit{T} means the speaker's third visit.

\subsubsection{Named structure}
Here is an example of speaker files.

\noindent
\texttt{
 T0003\// \\
	\hspace*{1em} 003MIC\//  \\
	\hspace*{1em} 003I0.25M\//  \\
	\hspace*{1em} 003PCM5M\//  \\
	\hspace*{1em} 003PCML3M\//  \\
	\hspace*{1em} 003PCM3M\//  \\
		\hspace*{2em} ...... \\
		\hspace*{2em} T0003\_003PCM3M\_recorded14\_0308\_normal.wav \\
		\hspace*{2em} ......
	}	

\noindent The corresponding structure is as follows, \\

$<$visit$><$spk\_id$>$\// \\
\hspace*{4em}$<$spk\_id$><$device\_and\_distance$>$\// \\
\hspace*{6em} $<$visit$><$spk\_id$>$\_$<$spk\_id$><$device\_distance$>$\\ \rightline{\_$<$channel\_id$>$\_$<$utt\_id$>$\_$<$speed$>$.wav} \\

In PCM (microphone array), \texttt{recorded 2} stand for channel 0, \texttt{recorded 6} denotes channel 4 and so on (in total there are 18 channels in each PCM, \texttt{recorded 0,1} is empty).

Here we provide three examples.

\begin{itemize}

	\item \texttt{F0148\_148I0.25M\_1\_0218\_normal.wav} means this audio is the utterance 218 in the first visit of speaker 148, the recorded device is iPhone at a distance of 25cm. \texttt{1} stands for channel \texttt{1}, which is meaningless since iPhone at 25cm distance only contains one channel.
	
	\item \texttt{S0183\_183MIC\_Tr2\_0138\_normal.wav} means this audio is the utterance 138 in the second visit of speaker 183, the recorded device is a close-talking MIC, \textit{Tr2} stands for close-talking. In this challenge, we just provide one close-talking microphone.

	\item \texttt{T0003\_003PCML3M\_recorded14\_0308\_normal.wav} means this audio is the utterance 308 in the third visit of speaker 3, the recorded device is a microphone array (this audio is channel 12 in this array), located at 3m distance in front of the speaker, on the left side with a 30 degrees angle.

\end{itemize}

In this challenge database, we provide three randomly selected microphone arrays out of the total six arrays in the training and development set; for each microphone array, we only provide 4 channels' data (channel 0,4,8,12, denoted by recorded 2,6,10,14) due to the large size of the whole database.

\subsection{The SLR-85 HI-MIA database}
The original HI-MIA database includes two sub databases, which are the AISHELL- wakeup1 with 254 speakers and the AISHELL-2019B-eval with 86 speakers. The content of utterances covers two wake-up words, \textit{`ni hao, mi ya'} in chinese and \textit{`Hi, Mia'} in English.

During the recording process, seven recording devices (one close-talking microphone and six 16-channel circular microphone arrays) were set in a real smart home environment. The 16-channel circular microphone array records signals in the 16kHz, 16 bit format, and the close-talking microphone records waveforms in the 44.1kHz, 16 bit format.

\footnote{\texttt{http://openslr.org/85/}}The SLR-85 HI-MIA open source database is the 2019 AISHELL Speaker Verification Challenge database which is a subset of the original HI-MIA database.
It contains one close-talking microphone and three microphone arrays located at 1m, 3m and 5m distance right in front of speaker. The SLR-85 HI-MIA database covers all the 254 speakers, but only includes the mandarin utterances.

For more details, please refer to the latest version (v3) of \cite{hi-mia}.

\section{Task Description}
\subsection{Task 1:Far-Field Text-Dependent Speaker Verification from single microphone array}

\subsubsection{Training data}

The training data includes 120 speakers and each speaker has 3 visits. In each visit, there are multiple (\textit{`ni hao, mi ya'}) text-dependent utterances as well as multiple text-independent utterances. The recording from five recording devices for each utterance are provided for training. These five recording devices include one close-talk microphone, one 25cm distance cellphone, and three randomly selected microphone arrays (4 channels per array).

Any publicly open and freely accessible speech database shared on openslr.org before Feb 1st 2020 (including SLR-85 HI-MIA) can be used in this challenge.
\subsubsection{Development Data}

The Development data includes 35 speakers and each speaker has 3 visits. In each visit, there are multiple (\textit{`ni hao, mi ya'}) text-dependent utterances as well as multiple text-independent utterances. The recording from five recording devices for each utterance are provided. These five recording devices include one close-talk microphone, one 25cm distance cellphone, and three randomly selected microphone arrays (4 channels per array).

\subsubsection{Evaluation Data}

The evaluation data includes 80 speakers and each speaker has 3 visits. In each visit, there are multiple (\textit{`ni hao, mi ya'}) utterances, The recording from two recording devices for each utterance are provided. These two recording devices include one 25cm distance cellphone and one randomly selected microphone arrays (4 channels per array).

The recording from 25cm distance cellphone will be selected as enrollment and recording from single far-field microphone array will be used for test. For any true trial, the enrollment and the testing utterances are from different visits of the same speaker.

\subsection{Task 2: Far-Field Text-Independent Speaker Verification from single microphone array}

\subsubsection{Training data}

The same as the training data for task 1.

\subsubsection{Development Data}

The same as the development data for task 1.

\subsubsection{Evaluation Data}

The evaluation data includes 80 speakers and each speaker has 3 visits. In each visit, there are multiple text-independent utterances, The recording from two recording devices for each utterance are provided. These two recording devices include one 25cm distance cellphone and one randomly selected microphone arrays (4 channels per array).

The recording from 25cm distance cellphone will be selected as enrollment and recording from single far-field microphone array will be used for test. For any true trial, the enrollment and the testing utterances are from different visits of the same speaker.

\subsection{Task 3: Far-Field Text-Dependent Speaker Verification from distributed microphone arrays}

\subsubsection{Training data}

The same as the training data for task 1.

\subsubsection{Development Data}

The same as the development data for task 1.

\subsubsection{Evaluation Data}

The evaluation data includes 80 speakers and each speaker has 3 visits. In each visit, there are multiple (\textit{`ni hao, mi ya'}) utterances. For each utterance, its corresponding recordings from one 25cm distance cellphone and 2-4 randomly selected microphone arrays are provided. For each microphone array, the selected four microphones are equally distributed along the circle with a random start channel index to simulate the scenarios with unknown array orientation angles. (e.g. channel 0, 4, 8, 12; channel 1, 5, 9, 13; channel 6, 10, 14, 2, etc.)

Recording from 25cm distance cellphone will be selected as enrollment and the recordings from 2-4 randomly selected far-field microphone arrays will be used for test. For any true trial, the enrollment and the testing utterances are from different visits of the same speaker.

There is no overlapping among the speakers in the training data, development data, evaluation data in task 1, task 2, and task 3.

\section{Evaluation Rules}

\subsection{Evaluation Results}
Before the mid-term deadline of score submission, each team have 5 times to submit the result. We sincerely suggest each team to test the system performance on the development set due to limited opportunities for submission. The evaluation set and the development set are from the same database, the only difference is that the development set only has 35 speakers, while the evaluation set for each task  has 80 speakers.

After the mid-term deadline and before the final deadline, each team has another 5 times to submit the final score file.

For the results on the evaluation set, we will release the results calculated based on a fixed 30\% of the trials in the leaderboard. So ranking on the leaderboard is not the final ranking. We will announce the official result in the Interspeech 2020 FFSVC special session. We encourage each team to explore more novel ideas, not just for the first place.

\subsection{The trials}
The trial file consists of three segments: enrollment audio ID, test audio ID and label. Label denotes that the trial is target or non-target. The 25cm distance iPhone signal is selected as the enrollment data, and the microphone array audio is considered as the testing data.

\subsection{Performance Measures}
In this challenge, we will use several metric to evaluate the system performance. The primary metric we adopt is the min $C_{det}$ cost value. In addition, Equal Error Rate (EER) and C$_{llr}$  will be provided to participant as auxiliary metrics.
\subsubsection{Primary measures}
The primary metric is based on the following detection cost function which is the same function as used in the
NIST 2010 SRE, but with modified parameters \cite{voice_plan}. It is a weighted
sum of miss and false alarm error probabilities in the form:

\begin{equation}
C_{det}=C_{miss}\times P_{miss}\times P_{tar} + C_{fa}\times P_{fa}\times (1-P_{tar})
\end{equation}

We assume a prior target probability, P$_{tar}$ of 0.01 and equal costs between misses and false alarms. The model parameters are 1.0 for both C$_{miss}$ and C$_{fa}$.
The C$_{det}$ will be normalized by P$_{tar}$ the same way as in \cite{voice_plan}.

We use the minimum C$_{det}$ as our primary metric.
\subsubsection{Alternate Performance Metrics}
The EER and C$_{llr}$ will be provided as alternate performance metrics and  a brief description of C$_{llr}$ is provided as follows:

To analyze how well a system performs and is calibrated across different operating points, a log-likelihood ratio based cost metric, C$_{llr}$, is suggested. Assuming trials scores are represented as LLRs, then C$_{llr}$ can be calculated as \cite{cllr},
\begin{equation}
C_{llr}=\frac{1}{2\times \log(2)}\times \Bigg(\frac{\sum \log(1+1/s)}{N_{tar}} + \frac{\sum \log(1+s)}{N_{non}}\Bigg)
\end{equation}
where $s$ is the likelihood ratio for a trial, and N$_{tar}$ and N$_{non}$ represent the number of target and non-target trials, respectively.

\section{Registration and Submission}

\subsection{Registration}
First, please create an account if you do not have one. We kindly request you to associate your account to an institutional e-mail. The organizing committee reserves the right to revoke your access to the challenge sites otherwise, please read the evaluation plan carefully. If you are part of a team, at least one person in your team will need an account to participate. Make sure to set the name of your team in the user's profile, or it will not be visible on the leaderboard.

Participants can register in one or more tasks. If your team participates in multiple tasks, we kindly request you to use the same user account to participate in all tasks.

Please note that any deliberate attempts to bypass the submission limit (for instance, by creating multiple accounts and using them to submit) will lead to automatic disqualification.

In case of any issues, the final interpretation right belongs to the organizing committee.
\subsection{Submission}

\subsubsection{Score submission}
Participants are required to submit at least one valid score file for each participating task to the FFSVC 2020 platform. The score files need to follow the following rules:
\\
\\
$<$TeamName$>$\_$<$Task$>$\_$<$SystemNumber$>$.txt\\

The score files should be in UTF-8 format with one line per trial. Each line must include two space-delimited fields: \\
\\$<$enrol\_data$><$space$><$test\_data$><$space$><$score$>$ \\

We will provided a sample in the challenge website.

\subsubsection{System description submission}
Each registered team is required to submit a technical system description report. Please submit this report using the Interspeech 2020 paper template. All reports must be a minimum of 2 pages (including references). Reports must be written in English. The system description does not need to repeat the content of the evaluation plan, such as the introduction of database, evaluation metric, etc. The system description must include the following items:
\begin{itemize}
	\item a complete description of the system components, including front-end (e.g., speech activity detection, features, normalization, front-end speech enhancement) and back-end (e.g., background models, i-vector/embedding extractor, Probabilistic Linear Discriminant Analysis (PLDA), speaker features fusion) modules along with their configurations (i.e., filterbank configuration, dimensionality and type of the acoustic feature parameters, as well as the acoustic model and the backend model configurations).
	
	\item a complete description of the data partitions used to train the various models.
	
	\item performance of the submission systems on the development dataset (calculated based on the provided tools) and the evaluation dataset (calculated by the challenge platform). Teams are encouraged to quantify the contribution of their major system components that they believe resulted in significant performance gains.\footnote{\texttt{https://www.nist.gov/system/files/documents/2019/\\07/22/2019\_nist\_speaker\_recognition\_challenge\_v8.pdf}}
	
	\item novel ideas, strategies and methods are strongly recommended to be shared.
	
	\item a report of the model size,  CPU (single threaded) and GPU execution times as well as the amount of memory used to process a single trial (i.e., the time and memory used for creating a speaker model from enrollment data as well as processing a test segment to compute the score).

\end{itemize}

\subsubsection{Paper submission}
The organizing committee highly encourages the participating teams to submit a paper to the INTERSPEECH 2020 Far-Field Speaker Verification Challenge special session. However, paper contributions within the scope are also welcome if the authors do not intend to participate in the Challenge itself. In any case, please submit your paper until 30 March 2020 (and final results by 15 June 2020) using the standard style info and respecting length limits, and submit to the Interspeech 2020 paper submission system. Important: as topic you should choose only this Special Session (FFSVC 2020). The papers will undergo the normal review process similar to the regular session papers.

\section{Schedule}\label{sec:conclusions}
\begin{itemize}
\item{Feb 1st: Releasing the training and development data as well as the evaluation plan}

\item{March 1st: Releasing the evaluation data and launching the leaderboard (30\% of the trials)}

\item{March 15th: Challenge registration deadline}

\item{March 23th: Mid-term deadline of score submission (up to 5 chances)}

\item{March 30th: Interspeech 2020 paper submission deadline}

\item{June 15th: Final deadline of score submission (another 5 chances)}

\item{Interspeech 2020 Camera Ready Paper deadline: System description submission deadline}

\item{Interspeech 2020 special session: Official results announcement}
\end{itemize}

\bibliographystyle{IEEEtran}
\bibliography{arxiv}

\end{document}